\documentclass[twocolumn,english]{revtex4}
\usepackage[T1]{fontenc}
\usepackage[latin9]{inputenc}
\usepackage{amsmath}
\usepackage{amssymb}
\usepackage{graphicx}

\makeatletter
\@ifundefined{textcolor}{}
{%
 \definecolor{BLACK}{gray}{0}
 \definecolor{WHITE}{gray}{1}
 \definecolor{RED}{rgb}{1,0,0}
 \definecolor{GREEN}{rgb}{0,1,0}
 \definecolor{BLUE}{rgb}{0,0,1}
 \definecolor{CYAN}{cmyk}{1,0,0,0}
 \definecolor{MAGENTA}{cmyk}{0,1,0,0}
 \definecolor{YELLOW}{cmyk}{0,0,1,0}
 }

\makeatother

\usepackage{babel}
\begin{document}

\title{Detecting paired and counterflow superfluidity via dipole oscillations}

\author{Anzi Hu$^{1}$, L. Mathey$^{1,2}$, Eite Tiesinga$^{1}$, Ippei Danshita$^{3}$,
Carl J. Williams$^{1}$ and Charles W. Clark$^{1}$}

\affiliation{$^{1}$Joint Quantum Institute,University of Maryland and National
Institute of Standards and Technology,Gaithersburg, MD 20899\\
 $^{2}$Zentrum f�r Optische Quantentechnologien and Institut f�r
Laserphysik, Universit�t Hamburg, 22761 Hamburg, Germany\\
 $^{3}$Computational Condensed Matter Laboratory, RIKEN, Wako, Saitama
351-0198, Japan}
\begin{abstract}
We suggest an experimentally feasible procedure to observe counterflow
and paired superfluidity in ultra-cold atom systems. We study the
time evolution of one-dimensional mixtures of bosonic atoms in an
optical lattice following an abrupt displacement of an additional
weak confining potential. We find that the dynamic responses of the
paired superfluid phase for attractive inter-species interactions
and the counterflow superfluid phase for repulsive interactions are
qualitatively distinct and reflect the quasi long-range order that
characterizes these phases. These findings suggest a clear experimental
procedure for their detection, and give an intuitive insight into
their dynamics. 
\end{abstract}
\maketitle
Rapid progress in experiments with ultra-cold atomic mixtures has
enabled the study of the rich quantum many-body phenomena of strongly
correlated multi-component systems in a controllable environment \cite{BBmixture_Inguscio}.
A central current objective of the ultra-cold atom community is the
realization and study of magnetic order, and, closely related, of
`$J^{2}/U$'--driven physics. $J$ refers to the tunneling energy,
and $U$ refers to the interaction strength in a Hubbard model, discussed
below. The most promising choice is to use bosonic atoms, due to the
higher degree of degeneracy that can be achieved in such systems.
In \cite{anzi} we identify the regime in which bosonic mixtures in
optical lattices sustain counterflow superfluidity (CFSF). Requiring
only repulsive contact interactions, this order is characterized by
particles binding with holes of the other species. Such particle-hole
pairing, or anti-pairing, leads to non-dissipative \emph{counter-flow}
of the two species while the net flow is zero \cite{Kuklov_CFSF}.
Besides CFSF, the ground state can also display paired superfluidity
(PSF), for attractive interactions \cite{KuklovDiagram,LM,Svistunov,anzi,anzinoise,Cirac_PSF,dembler_pd}.

We propose to realize CFSF. A key question is what maximal value of
$J^{2}/U$ can be used. CFSF competes energetically with the single-particle
superfluid (SF) phase. In \cite{anzi} we demonstrate that with $U_{12}/U\approx0.6\ldots0.8$,
CFSF is sustained up to $J/U\approx0.15$. $U_{12}$ is the interaction
strength between the two species, see below. When using, say, a mixture
of two hyperfine states of $^{87}$Rb in an optical lattice this can
be achieved by dislocating the two species slightly with a magnetic
field gradient \cite{spin_latt}.

In this paper we propose to use dissipationless counterflow as the
experimental signature of CFSF. In experiments of ultra-cold atoms
in optical lattices, transport properties are often studied by suddenly
displacing a confining harmonic potential and inducing dipole oscillations.
This type of experiment has been carried out to study superfluidity
of 1D \cite{trey_damp,1D_damp} and 3D \cite{3D_damp} Bose gases.
Here, we report the first many-body simulation to observe CFSF \cite{Kuklov_CFSF}.
We consider SF, PSF and CFSF states in a 1D Bose mixture in an optical
lattice confined in a weak trapping potential \cite{anzi,LM}. Using
the quasi-exact numerical method of time-evolving block decimation
(TEBD) \cite{Vidal}, we study the transport properties through the
dipole oscillations induced by either a brief or a constant displacement
of the harmonic trapping potential. Finding the qualitative features
described here experimentally, would demonstrate the existence of
CFSF order. We note that the same physical effects can also be achieved
by applying magnetic field gradients, and that similar results can
be expected for higher dimensions.

\begin{figure}
\includegraphics[width=7.5cm]{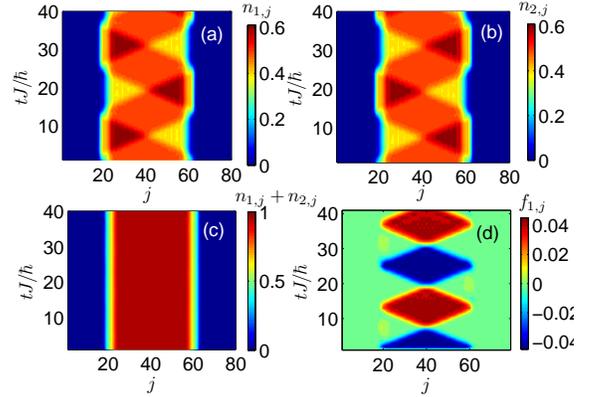}

\caption{\label{fig:1}(Color online) Dipole oscillations in the CFSF state
after a brief displacement of species 1, where $J/U=1/14,$ $U_{12}/U=0.6$,
$\Omega/U=10^{-3}$ and time $t$ is in units of $\hbar/J$. (a),
(b): density distributions of species 1 and 2, respectively. At $t=0$,
each species has a plateau at half-filling near the trap center. Then,
the trap of species 1 is perturbed by a brief displacement as shown
in Fig.~\ref{fig:2}(a). Species 1 and 2 are seen to begin moving
in opposite directions, whereas the total density is independent of
time as shown in panel (c). (d): local current of species 1, $f_{1,j}$,
defined in Eq. (\ref{eq:flux}). A positive current corresponds to
motion to the right. The diamond pattern in panel (d) reflects the
constant speed of flow. }
\end{figure}

We consider a 1D two-component Bose Hubbard model with a harmonic
trap centered at $c_{a}(t)$, 
\begin{eqnarray}
H & = & \sum_{a=1,2}\sum_{j}\left\{ -J(b_{a,j+1}^{\dagger}b_{a,j}+h.c.)+\Omega\left[j-c_{a}(t)\right]^{2}n_{a,j}\right.\nonumber \\
 &  & \left.+\frac{U}{2}n_{a,j}(n_{a,j}-1)\right\} +U_{12}\sum_{j}n_{1,j}n_{2,j}.\label{eq:hamiltonian}
\end{eqnarray}
 We denote the atomic species with index $a$, the lattice site with
index $j$, the number of lattice sites $N$, and we impose hard-wall
boundary conditions. In this paper, $N=80$. The species have the
same average filling factor, $\nu=M/N\leq1$, where $M$ is the number
of particles for each species. For all cases reported here, $M=20$.
We also assume that the repulsive intra-species interaction $U>0$,
hopping parameter $J>0$ and the spring constant $\Omega$ are the
same for both species. The inter-species interaction is given by $U_{12}$.
The operators $b_{a,j}^{\dagger}$ and $b_{a,j}$ are the creation
and annihilation operators for atoms of type $a$ on site $j$ and
$n_{a,j}=b_{a,j}^{\dagger}b_{a,j}$ is the atom number operator. We
assume that the trap centers, $c_{a}(t)$, are time dependent and
independently controllable. Their initial values are $c_{a}(t=0)=40.5$.
For convenience, we define the displacement $D_{a}(t)=c_{a}(t)-c_{a}(0).$
All distances are in units of the lattice constant $d$.

\begin{figure}
\includegraphics[width=7.5cm]{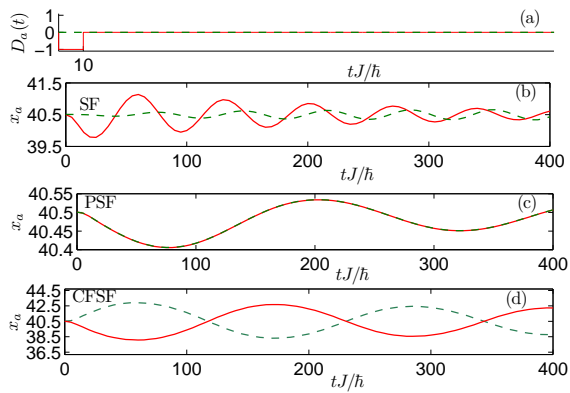}

\caption{\label{fig:2}(Color online) Center of mass oscillations of the two
species for SF, PSF, and CFSF states induced by a brief trap displacement
of species 1. (a): trap displacement vs. time. (b): evolution of the
SF state with parameters $J/U=1/8$, $U_{12}/U=-0.1$, $\Omega/U=2.5\times10^{-4}$.
The full (dashed) lines correspond to species 1 (2). (c): PSF state
for $J/U=1/8$, $U_{12}/U=-0.7$, $\Omega/U=2.5\times10^{-5}$. The
species oscillate in phase as a result of their pairing order. (d):
CFSF state for $J/U=1/8$, $U_{12}/U=0.7$ and $\Omega/U=10^{-3}$.
The species oscillate out of phase due to the anti-pairing order. }
\end{figure}

\begin{figure}
\includegraphics[width=8cm]{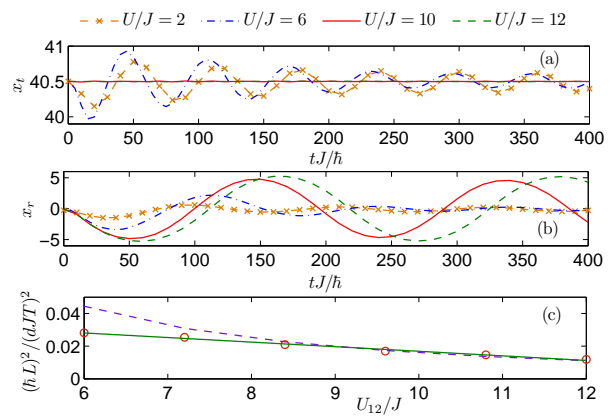}

\caption{\label{fig:3}(Color online) Dipole oscillation of the total and relative
center of mass; initial displacement as in Fig. \ref{fig:2}. (a):
$x_{\mathit{tot}}(t)$ for $U/J=2$, $6$, $10$, $12$; only in the
cases of $U/J=2$ and $6$ do we see significant induced oscillation.
(b): $x_{\mathit{rel}}(t)$ for the same $U/J$. A phase transition
from SF to CFSF is induced by changing $U/J$ while keeping $U_{12}/U$
and $\Omega/U$ fixed at $U_{12}/U=0.6$ and $\Omega/U=10^{-3}$.
The transition occurs for $U/J\approx8$. The transition is reflected
in the \emph{sudden} change in the behavior of the center of mass
motion. Panel (c) shows the ratio $(\hbar L)^{2}/(dJT)^{2}$ as a
function of $U_{12}/J$ in the CFSF region for fixed $U_{12}/U$,
where $T$ is the oscillation period and $L$ is the length of the
CFSF plateau extracted from the simulations. The markers are data
from our simulations. The full line is a linear fit and the dashed
line corresponds to Eq. (\ref{eq:relation2}). }
\end{figure}

The initial state is the ground state of $H$ at $t=0$ obtained by
the TEBD method with imaginary time propagation. The time evolution
is obtained with real-time TEBD propagation with the time step $\delta t$
equal to $\delta tJ/\hbar=0.05$, where $\hbar$ is the reduced Planck
constant. Previous works \cite{ippei_damp,Guido_damp} have applied
TEBD to simulate dipole oscillations in 1D single-species systems.
The simulations here use similar parameters as in Ref.\cite{ippei_damp}.
We analyze the time evolution of the system by studying the spatial
density distribution $\langle n_{a,j}(t)\rangle$, where the expectation
value is over the state of the system, and also the local flux or
current 
\begin{equation}
f_{a,j}=i(J/\hbar)\langle b_{a,j+1}^{\dagger}b_{a,j}-b_{a,j}^{\dagger}b_{a,j+1}\rangle,\label{eq:flux}
\end{equation}
 and the center of mass for each species 
\begin{eqnarray}
x_{a}(t) & = & \sum_{j}\frac{j}{M}\langle n_{a,j}(t)\rangle.\label{eq:CoM}
\end{eqnarray}
 We also define the total (relative) center of mass, $x_{\mathit{t}}(t)=\left[x_{1}(t)+x_{2}(t)\right]/2$
and $x_{\mathit{r}}(t)=x_{1}(t)-x_{2}(t)$.

We first consider the time evolution after a brief displacement of
the harmonic confinement of species 1. Figure \ref{fig:1} shows a
compelling example of the \emph{counter-flow} property of the CFSF
phase. Particle-hole pairing in the CFSF state requires that the particle
density of one species equals the hole density of the other: $\langle n_{j,1}\rangle=\langle1-n_{j,2}\rangle$.
For equal fillings, this means that the CFSF phase only occurs at
half-filling, $\nu=1/2$ \cite{anzi}. In a trapped system, the density
of the CFSF phase shows a plateau at half-filling near the center
of the trap. Near the edges, the system is SF. The impulse applied
to species 1 is generated by the brief displacement shown in Fig.~\ref{fig:2}(a).
It causes the species to move in opposite directions, as shown in
Fig.~\ref{fig:1}(a) and (b). The density of species $a$ at a given
position oscillates about $n_{a}=0.5$, and the density oscillations
of the two species differ in phase by $\pi$. The flux is reflected
at the steep SF edge of the atomic cloud. The total density distribution
stays unchanged and equals one (see Fig.~\ref{fig:1}(c)). Figure
\ref{fig:1}(d) shows the local current of species 1 as a function
of time. At any time, the flow occurs in only one direction. Moreover,
the color image shows characteristic diamond shapes. Within each diamond,
the current is nearly independent of lattice site and time, indicating
a constant velocity flow. This constant flow can be understood from
the constant total density of the CFSF state. As the energy due to
the trapping potential at each site only depends upon the total density
there, there is no potential energy cost in changing the local relative
density and the observed flow moves freely within the CFSF plateau.

In Fig.~\ref{fig:2}, we show an example of the center of mass motion,
$x_{a}(t)$, for SF, PSF, and CFSF initial states after applying a
small impulse to species 1. For the SF state, Fig.~\ref{fig:2}(b)
shows that only species 1 is excited immediately after the impulse.
Oscillatory motion of species $2$ is only induced gradually as a
result of the weak attraction between the two species. In Fig.~\ref{fig:2}(c)
we show the response of the PSF phase. Due to the pairing order, both
species move instantaneously and the time evolution of the center
of mass is identical. The impulse on species 1 is transformed into
a collective motion of both species and there is no relative motion.
In the CFSF phase shown in Fig.~\ref{fig:2}(d), we see that the
oscillatory motion of species $1$ is perfectly matched by an opposite
or counter-flowing motion of species $2$. We note that the sinusoidal
oscillation of the center of mass is not in contradiction with the
constant current flow shown in Fig.~\ref{fig:1}. Averaging over
lattice sites as in Eq. (\ref{eq:CoM}) erases this information.

\begin{figure}
\includegraphics[width=7.5cm]{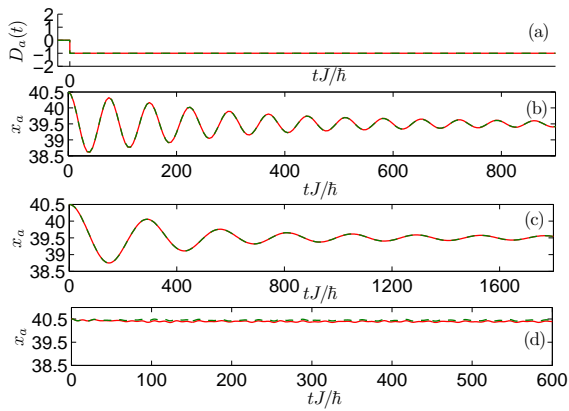}

\caption{\label{fig:4} Dipole oscillation of the center of mass of the two
species after a sudden displacement. (a): trap displacement vs. time.
(b), (c), (d): time evolution for the SF, PSF and CFSF states. The
system parameters are the same as in Fig.~\ref{fig:2} and the full
(dashed) lines correspond to species 1 (2). The depicted time scale
differs between panels. }
\end{figure}

In Fig.~\ref{fig:3} we plot the time evolution of the center of
mass of both species, $x_{\mathit{t}}(t)$, and the relative center
of mass, $x_{\mathit{r}}(t)$, after a brief displacement as in Fig.
\ref{fig:2} for several interaction strengths. We increase the intra-species
interaction $U$ from $2J$ to $20J$ and $U_{12}$ from $1.2J$ and
$12J$, with the fixed ratio $U_{12}/U=0.6$; we set $\Omega/U=10^{-3}$.
The system undergoes a phase transition from SF to CFSF around $U/J\approx8$.
In the SF state, the brief displacement of species 1 leads to damped
oscillations in both the relative and total center of mass. The maximum
amplitude of the oscillation increases with $U/J$. The damping rate
increases in this regime. As $U/J$ increases beyond the critical
point, the total center of mass motion is suddenly suppressed to almost
zero. On the other hand, the amplitude of the relative center of mass
keeps increasing through the phase transition. The damped oscillation
of the relative center of mass suddenly changes to nearly undamped
above the transition.

\begin{figure}
\includegraphics[width=8cm]{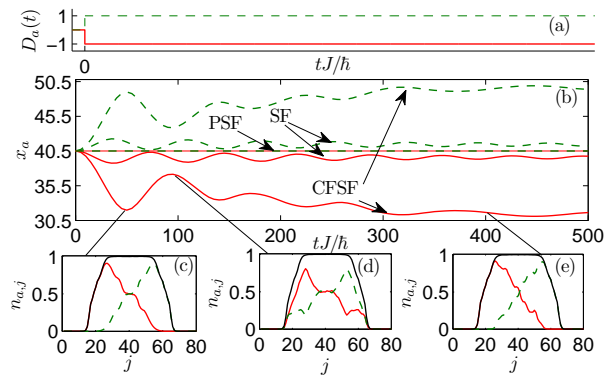}

\caption{\label{fig:5}Dipole oscillation of the center of mass after displacing
the two traps in opposite directions. (a): trap displacement vs. time.
(b): center of mass motion for the SF, PSF and CFSF states. The full
(dashed) line corresponds to species 1 (2). (c)-(e): density distributions
at $tJ/\hbar=55$, $100$ and $400$, respectively, for the CFSF state.
Parameters used for the three states are the same as in Fig. \ref{fig:2}.}
\end{figure}

We have also studied the dependence of the oscillation period on the
interaction strength in the CFSF state. As noted previously, the density
flow occurs at a constant velocity in this regime. The velocity can
be estimated with two methods: the Luttinger liquid theory and an
effective spin-1/2 model. Luttinger liquid theory predicts phonon-like
low energy excitations with a velocity 
\begin{equation}
v_{A}\sim\frac{2Jd}{\hbar}\sqrt{1-AU_{12}/J-BU_{12}/U},\label{eq:v_A}
\end{equation}
 where the non-universal coefficients $A$ and $B$ are assumed to
be constants. For a plateau of length $L$, the period of oscillation
is then $T=2L/v_{A}.$ For the analysis of our simulations, we find
it convenient to define the relationship based on Eq. (\ref{eq:v_A}),
\begin{equation}
\left(\frac{\hbar L}{dJT}\right)^{2}\sim1-AU_{12}/J-BU_{12}/U.\label{eq:relation}
\end{equation}

In the limit of large $U_{12}/J$ and $U/J$, we can derive another
relationship for the period. Assuming the particle-hole pairs are
hard-core bosons, Eq. (\ref{eq:hamiltonian}) can be mapped to an
effective spin-1/2 model \cite{dembler_pd}. Using linear spin-wave
theory\cite{linear_spin}, we obtain the phonon velocity $v_{A}$
as 
\begin{equation}
v_{A}=\frac{4J^{2}d}{U_{12}\hbar}\sqrt{1-U_{12}/U},\label{eq:v_A2}
\end{equation}
 and 
\begin{equation}
\left(\frac{\hbar L}{dJT}\right)^{2}=4(J/U_{12})^{2}(1-U_{12}/U).\label{eq:relation2}
\end{equation}
 Unlike for the Luttinger theory, this relationship does not have
free parameters. Figure \ref{fig:3}(c) shows the quantity $(\hbar L)^{2}/(dJT)^{2}$
as a function of $U_{12}/J$ for fixed $U_{12}/U$, where $L$ and
$T$ are obtained from the simulations. Overall, the quantity $(\hbar L)^{2}/(dJT)^{2}$
shows a linear trend consistent with the relationship in Eq. (\ref{eq:relation}).
The value is in good agreement with Eq. (\ref{eq:relation2}) especially
for $U_{12}/J>8.4$.

In Fig. \ref{fig:4} we consider the time evolution after a sudden
displacement for the SF, PSF and CFSF initial states. The displacement
is the same for both species. For the SF and PSF initial states, the
centers of mass of both species oscillate in phase around the new
minimum of the trap. The origin of the in-phase motion is different
for these two initial states. For the SF state, the species respond
independently to the same displacement. For the PSF state, the species
respond as pairs. The oscillation frequencies for the SF and PSF states
differ. This is not only because the states are confined by different
trapping potentials, but also because of the larger effective mass
of the pairs in the PSF state. The oscillations exhibit damping that
can be attributed to strong quantum fluctuations in 1D systems \cite{Guido_damp,ippei_damp,trey_damp}.
For the CFSF state, the motion is overdamped and the center of mass
remains at the original equilibrium position. The small oscillations
that are visible in the time evolution of $x_{1}(t)$ and $x_{2}(t)$
are due to oscillations of the superfluid at the edges of the atomic
cloud. For this type of displacement, the impact on both species is
the same and it only affects the total density. Because the motion
of the total density is suppressed for the CFSF state, the displacement
can not induce a response.

In Fig. \ref{fig:5} we consider a sudden displacement of the two
traps in opposite directions. For the SF state, the two species oscillate
around their respective new trap minima. The PSF state has no response,
because the pairing order resists the force imparted by the trap displacement.
The most dramatic response is in the CFSF state. Because the displaced
traps act as an effective linear potential, $2\Omega(D_{1}+D_{2})j$,
on the relative density, the two species are driven to move apart,
until they reach the edge of the CFSF plateau. \ref{fig:5}(c)-(e)
show density profiles at different times during the time-evolution.

By increasing the displacement amplitude, we can potentially break
the (anti-)pairing order. We can break the PSF pairing by applying
a large opposite displacement, while the CFSF pairing can be broken
by a large identical displacement. The displacement amplitude then
becomes an indicator of the binding energy of the pairs in the PSF
state and the anti-pairs in the CFSF state.

In conclusion, we have studied the dipole oscillation of 1D two-component
Bose mixtures in an optical lattice and a weak confining harmonic
potential. The oscillation is induced by displacing the harmonic confinement
either briefly or suddenly. We have shown that the response of the
system shows much richer features than its single-component counterpart.
In the two-component system, there are three long-range orders: the
superfluid, paired superfluid and counter-flow superfluid orders.
For the PSF and CFSF states, the suppression of individuality leads
to distinct dipole oscillations dependent on the character of the
perturbation. We have shown that this is the consequence of pairing
and anti-pairing ordering for the two states. For the CFSF state,
which forms plateaus of constant total density, the dipole oscillations
resemble the oscillation of a confined homogeneous system with free
propagation between boundaries.

We acknowledge helpful discussions with Trey Porto, Steven Olmschenk,
Dominik Schneble and Yoshiro Takahashi. This work was supported by
the U.~S.~Army Research Office under the Atomtronics MURI Program.
I.D. is supported by KAKENHI (22840051) from JSPS. The computational
work reported here was partially done on the RIKEN Cluster of Clusters
facility.

\end{document}